# The role of metacognition in troubleshooting: an example from electronics

Kevin L. Van De Bogart[1], Dimitri R. Dounas-Frazer[2],
H. J. Lewandowski[2,3], and MacKenzie R. Stetzer[1]

[1]Department of Physics and Astronomy, University of Maine, 120 Bennett Hall, Orono, ME, 04469-5709
[2]Department of Physics, University of Colorado Boulder, 390 UCB, Boulder, CO, 80309
[3]JILA, University of Colorado Boulder, 440 UCB Boulder, CO 80309

**Abstract.** Students in physics laboratory courses, particularly at the upper division, are often expected to engage in troubleshooting. Although there are numerous ways in which students may proceed when diagnosing a problem, not all approaches are equivalent in terms of providing meaningful insight. It is reasonable to believe that metacognition, by assisting students in making informed decisions, is an integral component of effective troubleshooting. We report on an investigation of authentic student troubleshooting in the context of junior-level electronics courses at two institutions. Think-aloud interviews were conducted with pairs of students as they attempted to repair a malfunctioning operational-amplifier circuit. Video data from the interviews have been analyzed to examine the relationship between each group's troubleshooting activities and instances of socially mediated metacognition. We present an analysis of a short episode from one interview.

## I. INTRODUCTION

Students are typically required to take several laboratory courses as part of an undergraduate physics program. To date, however, relatively little research has focused on students' activities within the instructional laboratory environment. It is therefore not surprising that laboratories were identified by the 2012 report on discipline-based education research as a critical area of interest [1]. While there are ongoing efforts to transform upper-division laboratory instruction [2], such efforts further underscore the need for targeted investigations related to laboratory learning goals and practices. In particular, student ability to troubleshoot was identified as an important learning outcome in the recently endorsed AAPT Recommendations for the Undergraduate Physics Laboratory Curriculum [3]. The report suggests that students should be able to approach troubleshooting in an iterative and logical way by the completion of their physics degree. Indeed, students are frequently expected to engage in troubleshooting when working with complex systems in upper-division courses such as electronics.

While there has been considerable research in PER on student understanding of introductory circuits [7,8], much less work has been conducted in the context of upper-division electronics. Moreover, much of this upper-division work has primarily focused on student learning of specific topics including operational-amplifier (op-amp) circuits [9,10], phase relationships in AC circuits [11], and RC filters [12]. None of this work has expressly targeted laboratory skills such as troubleshooting.

We are currently investigating student troubleshooting in upper-division electronics courses through the use of two complementary theoretical frameworks: the Socially Mediated Metacognition Framework (SMMF) [4] and the Experimental Modeling Framework (EMF) [5]. This article focuses on preliminary results from an analysis using the perspective of metacognition, while a companion article presents results emerging from the EMF.

In accordance with the work of Schaafstal *et al.* [6], we use the term troubleshooting to refer to a comprehensive process that includes identifying the existence and nature of a problem and taking corrective action. Their study of navy technicians found that a standard course on content understanding was insufficient to prepare students to troubleshoot radar systems, but that supplementary training in structured troubleshooting resulted in significant improvements.

The term *metacognition* refers broadly to thinking about one's own thinking and is often subdivided into categories of self-assessment (*e.g.,* understanding and communicating one's own thought processes), self-regulation (*e.g.,* consideration of how to perform long tasks), and knowledge from previous experience [13]. Schoenfeld's work on mathematical problem solving highlights the relevance of the self-regulation aspect of metacognition. He found in interviews that students frequently selected a particular solution method and set about implementing it without further consideration of either the appropriateness of their approach or other alternatives [14]. In contrast, an experienced mathematician spent the majority of his time analyzing, planning, and then assessing the utility of specific actions rather than implementing such approaches immediately. Students explicitly taught by Schoenfeld to adopt specific metacognitive practices were similarly found to exhibit more expert-like problem-solving behavior. Indeed, metacognition is believed to assist in the selection of productive problem-solving approaches via ongoing assessment [15]. This paper aims to demonstrate that metacognition may play a similar role in troubleshooting in an upper-division laboratory context.

Due to the collaborative nature of student work in many upper-division laboratory courses, we adopt a framework of socially mediated metacognition, or SMMF [4]. A similar approach was used in the work of Lippmann Kung and Linder to examine the impact of laboratory instruction on student metacognitive behavior in the context of an introductory physics course [16]. This framework leverages the fact that metacognition at the group level is necessarily expressed through collaborative interactions. As researchers, we can categorize these interactions in terms of both the metacognitive content individuals are contributing to the group and the conversational transactives (*i.e.* verbal exchanges) that serve to aid the group members in understanding one another.

In this paper, we describe preliminary findings from an ongoing investigation of how students troubleshoot an op-amp circuit similar to those covered in a junior-level electronics course. To this end, we have selected a short vignette from one task-based interview and coded it for instances of socially mediated metacognition to illustrate the framework's applicability to troubleshooting circuits. We demonstrate that the coding scheme adequately captures students' metacognitive interactions and provides evidence that metacognition can be an asset when troubleshooting circuits.

## II. INTERVIEW TASK

*Overview.* A total of 8 task-based interviews were performed, 4 conducted concurrently with the junior-level electronics course at the University of Colorado Boulder and 4 conducted one semester after a similar course at the University of Maine. Interviews were performed with pairs of students (all of whom were used to working in pairs in lab) and lasted from 25 minutes to one hour. For the interviews, an op-amp circuit containing two amplification stages was set up ahead of time. Students were given a diagram of the circuit (Fig. 1) and were told to approach the circuit as though it had been built by their peers, was found to be malfunctioning, and had not yet been successfully diagnosed. Interview participants were tasked with identifying all problems and repairing the circuit, all with only minimal interference from the interviewer.

*Circuit functionality.* The cascading amplifier circuit used for the interviews has two distinct stages. The op-amp on the left, along with $R_1$ and $R_2$, form stage 1, which is a non-inverting amplifier with an output voltage that is twice as large as the input voltage (*i.e.,* a gain of 2). $R_3$, $R_4$, and the rightmost op-amp comprise stage 2, which is an inverting amplifier with a gain of −10; note that the negative sign indicates that the output voltage of stage 2 is 180° out of phase with its input voltage. The gains from the two stages may be multiplied together to obtain the output voltage of the entire circuit, $V_{out}$, which is twenty

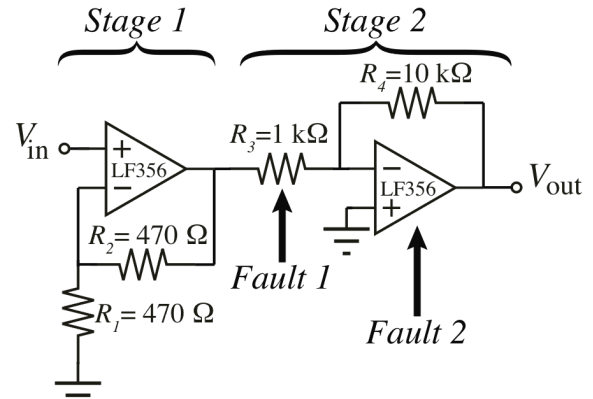

**Figure 1**. The cascading amplifier circuit. The diagram given to students did not identify the stages or faults.

times larger than, and 180° out of phase with, $V_{in}$, the input voltage of the circuit.

The physical circuit given to students was constructed with two distinct, intentional flaws. The first was that the resistor $R_3$ was 100 Ω instead of the nominal value of 1 kΩ (as indicated on the diagram). This resulted in stage 2 having a gain ten times larger than expected. The second was that the op-amp in stage 2 was damaged in such a way that the output was constant at approximately −15V, regardless of input.

This combination of errors was specifically chosen for two reasons. First, neither impacted the operation of stage 1. This allowed for the use of a split-half troubleshooting strategy, in which half of the circuit can be identified as working and thus the problem space can be reduced [17]. Second, the presence of two distinct errors required students to engage in multiple rounds of troubleshooting. We anticipated that students would typically identify the incorrect resistor first, as it differed visually from the correct resistor, but would struggle to identify the second fault.

*Illustrative episode.* In this article, we analyze an episode that is approximately 90 seconds in length and took place roughly halfway though the interview. This particular episode was chosen because it demonstrates the utility of both frameworks (SMMF and EMF) for interpreting troubleshooting incidents. Moreover, it highlights the complementary nature of these frameworks in providing a more complete and coherent understanding of troubleshooting.

Earlier in the interview, the two students in this episode had oriented themselves to the circuit, connected their power sources and measurement devices, and calculated the expected output of the entire circuit. It is important to note that the students had set their oscilloscope to ac coupling, which removed the dc portion of the signals and thus made the −15 V output of the circuit appear to be a very small ac signal centered about 0 V.

## III. CODING METHODOLOGY

In order to analyze our interviews, we employed the two-level coding scheme developed by Goos *et al.* [4] First, transcripts were coded line by line for their metacognitive function, defined as follows:

- *New ideas*, which include new information and new approaches being verbalized, and
- *Assessment* of information and approaches, including the appropriateness of a *strategy*, the sensibility of *results*, and students' own *understanding*.

To capture the interactions between participants, transcripts were also coded according to the following three transactive exchanges:

- *Self-disclosure,* in which an attempt is made to clarify one's thinking to a partner,
- *Feedback requests*, in which one invites a partner to critique one's thinking, and
- *Other-monitoring,* in which one's statements and questions are used in an attempt to better understand a partner's thinking.

## IV. EPISODE TRANSCRIPT

1   **S1:** So, it's doubled $V_{in}$, but it's not inverted it.
2        And it shouldn't be inver- should be-
3   **S2:** Is that an inverting amplifier?
4   **S1:** No, it's not. An inverting amplifier is
5        connected to the- $V_{in}$ is connected to the
6        negative terminal, right?
7   **S2:** Yeah, yeah.
8   **S1:** So it shouldn't be inverted.
9        So this one- [Points to schematic]
10  **S2:** Well, neither of them are inverting.
11  **S1:** Oh, yes. [Points to schematic]
12       This one is inverting.
13  **S1**: The second one is inverting
14  **S2:** But our $V_{out}$ right now isn't inverting. [Points to oscilloscope]
15       So that probably means that these positive,
16       plus and minus terminals on the second one
17       are just mixed up. [Points to schematic]
18  **S1:** Why?
19  **S2:** Because it's not inverting.
20       So this is an inverting amplifier, so they just
21       mixed up the plus and minus [Points to schematic]
22  **S1:** But this one's not doing anything at all. [Points to schematic]
23       The way this is drawn here is inverting.
24  **S2:** Yeah. But on here- [Points to circuit]
25  **S1:** On here it's not- [Leans over circuit]
26       there's no output at all.
27       I mean there's this tiny- [Points to oscilloscope]
28  **S2:** What do you mean? [Points to oscilloscope]
29       Yeah, there's-
30  **S1:** I guess, but that's like-
31  **S2:** How much- Well, how big is it?
32  **S1:** It's tiny. It's like ten millivolts.
33  **S2:** Oh. Well, okay.
34       We have a good output for the first op-amp,
35       so we are going to have- the problem is in the
36       second one.

## V. ANALYSIS: LINES 1-8

In this portion of the episode, S2 helps regulate how S1, and thus the pair, is thinking about the circuit. S1 begins to discuss what he thinks the circuit should do, but he is not sure what the behavior should be (1-2: *assessment of results, self-disclosure*). S2 then intercedes with his question, "Is that [stage 1] an inverting amplifier?" (3: *other-monitoring*). S1 concludes that it is not, and articulates his idea of what characterizes an inverting amplifier (4-6: *new idea, feedback request*). Finally, S2 briefly confirms that he is correct, and his partner finishes his assessment of what the first amplifier should do (7-9: *assessment of results, self-disclosure*).

This episode demonstrates a way in which socially mediated metacognition may assist in troubleshooting through the assessment of the appropriateness of the pair's ideas. At the start of the episode, S1 does not have a clear prediction about whether or not the output of stage 1 of the circuit should be inverted. By considering what is required for a circuit to be an inverting amplifier and then concluding that this circuit does not meet those criteria, S1 has generated new information for the pair to work from and thus establishes more clear expectations for how the circuit should behave. When the EMF is applied to this same excerpt, we find that the process the students are undertaking may be placed into the larger structure of a cycle of model construction.

## VI. ANALYSIS: LINES 14-36

The second portion of this episode begins with S2 noting that the output is not inverted, and then putting forth the hypothesis that this is due to an accidental reversal of the two input connections (14-17: *new idea, self-disclosure*). His partner asks S2 why he thinks that is the case (18: *other-monitoring*), and S2 reiterates his idea. S1 responds by observing that the op-amp is not doing anything at all, and notes that there is only a tiny output signal (22-27: *assessment of results, new idea*). S2 wants to know what his partner means (28-29: *other-monitoring*) and asks for more details about the output (31: *new idea*). S1 responds that it is tiny– only 10 mV (32: *assessment of results*). Finally, S2 evaluates the state of their current understanding by noting that they know the first op-amp is good, and that something is wrong with the second one (33-36: *assessment of results, self-disclosure*).

In this instance, we see that there is a cyclic interaction between the two participants. S2 puts forth an idea, which S1 tries to further understand. S1 then puts forth his own idea implying that there is a more serious problem, and supports it with an observation. S2 attempts to get a better understanding of this idea, and seeks clarification via measurement to help support this idea. After being presented with evidence supporting S1's conclusion, S2 accepts that this is a problem, and then re-evaluates what they have learned from this interaction. This mutual attempt to understand each other's thinking is captured in our coding of back and forth *other-monitoring* exchanges.

Perhaps more importantly, this episode also demonstrates how metacognition may assist in troubleshooting by efficiently ruling out approaches that would not be productive. The idea that the output was unreasonably small overshadowed the original idea of switching inputs to correct the phase of the output voltage; the latter approach would not have repaired the circuit and would simply have been an extra experimental test. By considering the diagnostic power of an action before performing it, a student may compare that option to other alternatives and identify the best course of action in a timely fashion. In addition, minimizing the actions taken leaves fewer opportunities to inadvertently introduce additional errors into the circuit. From the modeling framework, this portion of the episode is demonstrative of the process that students use to assess proposals for revisions.

## VII. CONCLUSIONS

We have illustrated that the Socially Mediated Metacognition Framework is both suitable for and useful in the analysis of student troubleshooting of an electronic circuit. As we have shown, metacognitive regulation may assist students in solidifying their own understanding and in making strategic decisions when testing hypotheses. We note that this framework may not capture the more technical aspects of troubleshooting; such as taking measurements, making predictions, and constructing models. The Experimental Modeling Framework complements this analysis by providing further insight into both the role of model construction in troubleshooting and how students use models to make predictions, as shown in a companion paper [18]. In the future, we plan to analyze all interview data via both complementary frameworks in order to characterize student troubleshooting in a much more detailed way. The knowledge gained from this analysis will guide the development of instructional materials that support and promote effective student troubleshooting in electronics.

## ACKNOWLEDGEMENTS

Support by the National Science Foundation under Grant Nos. DUE-1323426, DUE-1245313, DUE-0962805, and DUE-1323101 is gratefully acknowledged.

[1] National Research Council, Discipline-Based Education Research: Understanding and Improving Learning in Undergraduate Science and Engineering, 2012.
[2] B. M. Zwickl, N. Finkelstein, and H. J. Lewandowski, Am. J. Phys. **81**, 63 (2013).
[3] J. Kozminski, H. J. Lewandowski, N. Beverly, S. Lindaas, D. Deardorff, A. Reagan, R. Dietz, R. Tagg, M. Eblen-Zayas, J. Williams, R. Hobbs, and B. M. Zwickl, AAPT Recommendations for the Undergraduate Physics Laboratory Curriculum, 2014 (unpublished).
[4] M. Goos, P. Galbraith, and P. Renshaw, Educ. Stud. Math. **49**, 193 (2002).
[5] B. M. Zwickl, D. Hu, N. Finkelstein, and H. J. Lewandowski, arXiv:1410.0881.
[6] A. Schaafstal, J. M. Schraagen, and M. van Berlo, Hum. Factors **42**, 75 (2000).
[7] L. C. McDermott and P. S. Shaffer, Am. J. Phys. **60**, 994 (1992).
[8] P. V. Engelhardt and R. J. Beichner, Am. J. Phys. **72**, 98 (2004).
[9] A. Mazzolini, T. Edwards, W. Rachinger, S. Nopparatjamjoras, and O. Shepherd, Latin-American J. Phys. Educ. **5**, 147 (2011).
[10] C. P. Papanikolaou, G. S. Tombras, K. L. Van De Bogart, and M. R. Stetzer, Am. J. Phys. (submitted).
[11] C. Kautz, in *Proc. World Eng. Educ. Flash Week,* Lisbon, 2011, edited by J. Bernardino and J. C. Quadrado (Higher Institute of Engineering in Lisbon, Lisbon, 2011), p. 228
[12] P. Coppens, M. De Cock, and C. Kautz, in *Proc. 40th SEFI Annu. Conf,* Leuven, 2012, edited by A. Avdelas (University of Leuven, Leuven, 2012), p196.
[13] J. Jacobs and S. Paris, Educ. Psychol. **22**, 255 (1987).
[14] A. Schoenfeld, in *Cognitive Science and Mathematics Education,* edited by A. Schoenfeld (Lawrence Erlbaum Associates, Hillsdale, 1987).
[15] T. van Gog, F. Paas, J. J. G. van Merriënboer, and P. Witte, J. Exp. Psychol. Appl. **11**, 237 (2005).
[16] R. Lippmann Kung and C. Linder, Metacognition Learn. **2**, 41 (2007).
[17] D. H. Jonassen and W. Hung, Educ. Psychol. Rev. **18**, 77 (2006).
[18] D. R. Dounas-Frazer, K. L. Van De Bogart, M. R. Stetzer, and H. J. Lewandowski in *Physics Education Research Conference Proceedings*, College Park, 2015, edited by A. D. Churukian, D. L. Jones, and L. Ding (submitted).